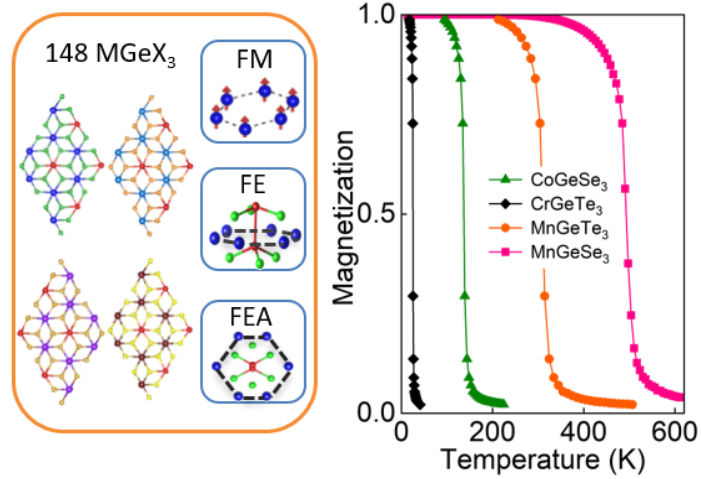

We systematically investigated the possible ferroic properties of 2D $MGeX_3$. Eight stable 2D ferromagnets, 21 2D antiferromagnets, and 11 stable 2D ferroelectric semiconductors including two multiferroic materials were predicted. $MnGeSe_3$ and $MnGeTe_3$ are predicted to be room-temperature 2D ferromagnetic half metals with $T_c$ of 490 K and 308 K, respectively.

# The atlas of ferroicity in two-dimensional MGeX$_3$ family: room-temperature ferromagnetic half metals and unexpected ferroelectricity and ferroelasticity

Kuan-Rong Hao[1], Xing-Yu Ma[1], Hou-Yi Lyu[1,3], Zhen-Gang Zhu[2,1], Qing-Bo Yan[3*], Gang Su[4, 1, 3*]

[1]School of Physical Sciences, University of Chinese Academy of Sciences, Beijing 100049, China.
[2]School of Electronic, Electrical and Communication Engineering, University of Chinese Academy of Sciences, Beijing, 100049, China.
[3]Center of Materials Science and Optoelectronics Engineering, College of Materials Science and Optoelectronic Technology, University of Chinese Academy of Sciences, Beijing 100049, China.
[4]Kavli Institute for Theoretical Sciences, and CAS Center of Excellence in Topological Quantum Computation, University of Chinese Academy of Sciences, Beijing 100190, China.
*Correspondence authors. Email: yan@ucas.ac.cn; gsu@ucas.ac.cn

**Two-dimensional (2D) ferromagnetic and ferroelectric materials attract unprecedented attention due to the spontaneous-symmetry-breaking induced novel properties and multifarious potential applications. Here we systematically investigate a large family (148) of 2D MGeX$_3$ (M = metal elements, X = O/S/Se/Te) by means of the high-throughput first-principles calculations, and focus on their possible ferroic properties including ferromagnetism, ferroelectricity, and ferroelasticity. We discover eight stable 2D ferromagnets including five semiconductors and three half-metals, 21 2D antiferromagnets, and 11 stable 2D ferroelectric semiconductors including two multiferroic materials. Particularly, MnGeSe$_3$ and MnGeTe$_3$ are predicted to be room-temperature 2D ferromagnetic half metals with $T_c$ of 490 and 308 K, respectively. It is probably for the first time that ferroelectricity is uncovered in 2D MGeX$_3$ family, which derives from the spontaneous symmetry breaking induced by unexpected displacements of Ge-Ge atomic pairs, and we also reveal that the electric polarizations are in proportion to the ratio of electronegativity of X and M atoms, and IVB group metal elements**

**are highly favored for 2D ferroelectricity. Magnetic tunnel junction and water-splitting photocatalyst based on 2D ferroic MGeX$_3$ are proposed as examples of wide potential applications. The atlas of ferroicity in 2D MGeX$_3$ materials will spur great interest in experimental studies and would lead to diverse applications.**

## 1. INTRODUCTION

The ferroic materials with ferromagnetic (FM) and ferroelectric (FE) orderings involve phase transitions with spontaneous symmetry breakings at critical temperature, thus they can maintain and transit between different phases under external stimuli, providing many possibilities for novel smart devices. Recently, novel 2D FM and FE materials have attracted great interest. Intrinsic ferromagnetism was observed in bilayer Cr$_2$Ge$_2$Te$_6$ [1] and monolayer CrI$_3$ [2] with Curie temperature $T_C$ = 28 K and 45 K, respectively. In addition, more 2D FM materials have been predicted, such as transition metal dichalcogenides (MX$_2$, M = metal atoms, X = S/Se/Te)), [3-6] transition metal halogenides (MX$_3$ and MX$_2$, X = Cl/Br/I) [7-10], transition metal oxides (MO, M$_2$O$_3$, MO$_2$) [11, 12], ternary transition metal compounds (MAX$_3$, A = Ge/Si/P/Sn, X = S/Se/Te) [13-20], etc. However, the currently available 2D FM materials with low Curie temperatures restrict their practical applications. Seeking for 2D ferromagnets with high $T_C$ especially above room temperature is thus highly needed. On the other hand, 2D In$_2$Se$_3$ [21], CuInP$_2$S$_6$ [22] and CuCrP$_2$S$_6$ [23] were shown to exhibit ferroelectricity, and a variety of 2D ferroelectric materials were also theoretically predicted, such as 2D MoS$_2$ [24], group-IV mono-chalcogenides (MX, M=Ge, Sn, X=S, Se) [25-28], CrN and CrB$_2$ [29], SbN and BiP [30], group-IV tellurides (XTe, X = Si, Ge, Sn) [31], ScCO$_2$ [32], M$_I$M$_{II}$P$_2$X$_6$ (M$_I$, M$_{II}$ = metal elements, X = O/S/Se/Te) [33-35], WTe$_2$ [36-38] and so on. Meanwhile, the ferroelasticity was predicted in phosphorene [25], group-VIB transition metal dichalcogenides (TMDs) [39] and several group-IV mono-chalcogenides [27], etc. Moreover, 2D multiferroic materials possessing simultaneously two or more intrinsic ferroic orders in one material have also been highly concerned [40-42]. Several 2D multiferroics such as CuCrP$_2$X$_6$ (X = S, Se) [23], CuCrX$_2$ (X = S, Se) [43], CrN and CrB$_2$ [29], group-IV chalcogenides [25, 27], ReWCl$_6$

[44] and $M_IM_{II}P_2X_6$ ($M_I$, $M_{II}$ = metal elements, X = O/S/Se/Te) [35] have been predicted. Those 2D ferroic materials would be promising candidates for miniaturizing functional devices such as spintronic transistors, and nonvolatile memory devices [45-50], etc.

Inspired by the experimental discovery of a 2D ferromagnet $Cr_2Ge_2Te_6$ [1], we here perform a systematic first-principles investigation on a large family of 2D $MGeX_3$ (M=metal atoms, X=O/S/Se/Te) focusing on their possible ferroic properties including ferromagnetism, ferroelectricity, and ferroelasticity, and obtain a comprehensive atlas of ferroicity. We discover eight stable 2D ferromagnets including five semiconductors and three half-metals, 21 2D antiferromagnetic (AFM) materials, and 11 stable 2D FE semiconductors and four 2D ferroelastic (FEA) materials, including two 2D multiferroic materials. Particularly, 2D $MnGeSe_3$ and $MnGeTe_3$ are predicted to be room-temperature FM half-metals with $T_C$ of 490 K and 308 K, respectively. It is revealed that IVB group elements are highly favored for the above discovered 2D ferroelectric $MGeX_3$ materials, and the physical origin of ferroelectricity is unveiled to be owing to the spontaneous symmetry breaking induced by unexpected vertical displacements of the Ge-Ge atomic pairs. Moreover, the applications of 2D FM half-metals for magnetic tunnel junctions and 2D FE materials in photocatalytic water-splitting are proposed. These findings highly enrich the family of 2D ferroic materials and would motivate great interest in exploring 2D $MGeX_3$ family as novel potential multi-functional materials.

**2. METHODS**

The first-principles calculations based on the density functional theory (DFT) are carried out with the Vienna *ab initio* simulation package (VASP) [51, 52]. The projector augmented wave method is used to describe the interaction between core and valence electrons [53]. The electron exchange-correlation functional is dealt with the generalized gradient approximation (GGA) in the PBE form [54]. The orbital-dependent on-site Coulomb interactions (U) have been considered with the values of 4.0, 2.0, 0.5 for *3d, 4d*, *5d* metal elements, respectively. The cutoff energy of plane-wave is taken as 500 eV and the total energy convergence threshold is $10^{-6}$ eV/atom.

The full structure optimizations on atomic positions and lattice vectors are performed until the maximum force on each atom was less than 0.001 eV/Å. To avoid the mirror interaction in the vertical direction, the vacuum space of 20 Å is introduced between adjacent mirror layers for simulating monolayer $MGeX_3$. The phonon dispersions are calculated to show the dynamical stability using the finite displacement approach in the PHONOPY package [55]. The hybrid functional HSE06 is further used to accurately calculate the electronic structures [56]. The climbing-image nudged elastic band method (CI-NEB) [57] is applied to determine the minimum energy path and barrier in the reversal process of electrical polarization. The optical adsorption spectra are obtained by employing the $G_0W_0$ approximation and Bethe-Salpeter equation (BSE) method [58, 59].

## 3. RESULTS AND DISCUSSIONS

**3.1 Geometric structures of the 2D $MGeX_3$ family**. Figure 1(a) and (b) indicate the schematic structures of the 2D $MGeX_3$ (M = metal elements; X = O/S/Se/Te). The blue, red and green balls represent metal M atoms, Ge and chalcogen X atoms, respectively. M atoms form a hexagonal honeycomb lattice and Ge-Ge pairs locate at the center of hexagons, which are linked by chalcogen atoms. The parallelogram indicates a primitive cell containing two formula units of $MGeX_3$. When Ge-Ge pairs are vertically bisected by the hexagonal plane of M atoms, the structure has an inversion symmetry and the corresponding space group is *P*-31*m* (No.162), as indicated in Fig. 1(b). Interestingly, two different types of geometric structures with spontaneous symmetry breaking are observed for some $MGeX_3$, which are illustrated in Figs. 1(c) and 1(d) as type-I and type-II, respectively. For type-I, the hexagonal lattice holds its original configuration while the Ge-Ge pair unexpectedly displaces along the *z* axis, which breaks the inversion symmetry and the space group reduces to *P*31*m* (157), corresponding to a "ferroelectric phase". For type-II, the Ge-Ge pairs are inclined to three different directions and three equivalent distorted structures *α*, *β* and *γ* are thus obtained with the space group altered from *P*-31*m* (No.162) to *Cm* (8), corresponding to a "ferroelastic phase". It is worth noting that the above two different types of spontaneous symmetry breakings and the corresponding ferroelectric and ferroelastic phases have not been reported for 2D MGeX$_3$ materials previously. Besides, to study the magnetic properties of all 2D MGeX$_3$, FM and several possible AFM spin configurations on honeycomb lattice are considered, as shown in Fig.1(e).

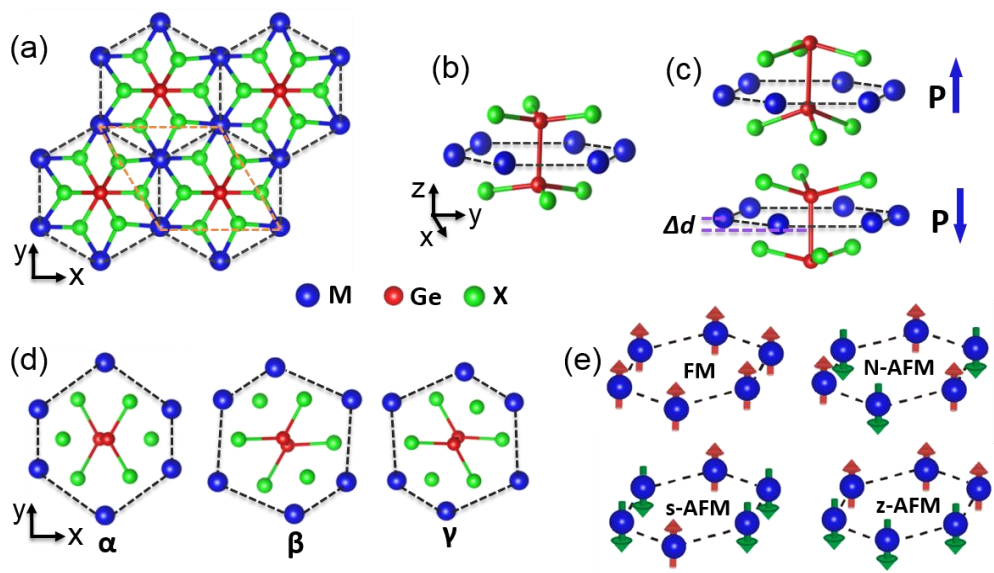

Figure 1. Schematic structures of 2D MGeX$_3$ (M = metal elements; X = O, S, Se, Te). The blue, red, and green balls represent M, Ge, and X atoms, respectively. (a) Top view and (b) side view; (c) type-I structures of ferroelectric phase with opposite out-of-plane polarization directions, $\Delta d$ indicates the off-centering displacement of Ge-Ge pair with respect to the hexagonal plane of M atoms; (d) three equivalently distorted type-II structures ($\alpha$, $\beta$, $\gamma$) of the ferroelastic phase. (e) The ferromagnetic (FM) and several antiferromagnetic (AFM) configurations on honeycomb lattice: Néel AFM (N-AFM), stripe AFM (s-AFM), and zigzag AFM (z-AFM).

**3.2 The workflow for searching ferroic 2D MGeX$_3$ materials.** Figure 2 illustrates the schematic flowchart of high-throughput first-principles investigations on geometric structures and ferroic properties of 2D MGeX$_3$ family. We designed 148 different 2D MGeX$_3$ by replacing M with 37 metal elements and replacing X with four chalcogen atoms (O, S, Se, and Te). For all of them, we performed full geometric relaxations in different magnetic configurations (indicated in Fig. 1(e)), and then obtained the optimized geometric structures with magnetic ground states, according to which each 2D MGeX$_3$ member can be labeled with ferromagnetic (FM), antiferromagnetic (AFM) and nonmagnetic (NM). On the other hand, based on the optimized geometric structures, their space groups and point groups can be obtained, which indicate the possible symmetry breakings and can be used to identify ferroelectric (FE) and ferroelastic (FEA) materials. For the so-obtained FM, FE and FEA materials, we examined their dynamical stabilities and then performed calculations of their electronic structures and other physical properties related to ferroicity. Finally, we discovered eight stable 2D ferromagnets including five semiconductors and three half-metals, 21 2D AFM materials, 11 stable 2D FE semiconductors and four stable 2D FEA materials, including two 2D multiferroic materials. The atlas of ferroicity in 2D MGeX$_3$ family is depicted in Fig. 3. Different colors indicate different ferroic properties of corresponding 2D MGeX$_3$, and two different colors in the same block implies a potential multiferroic material.

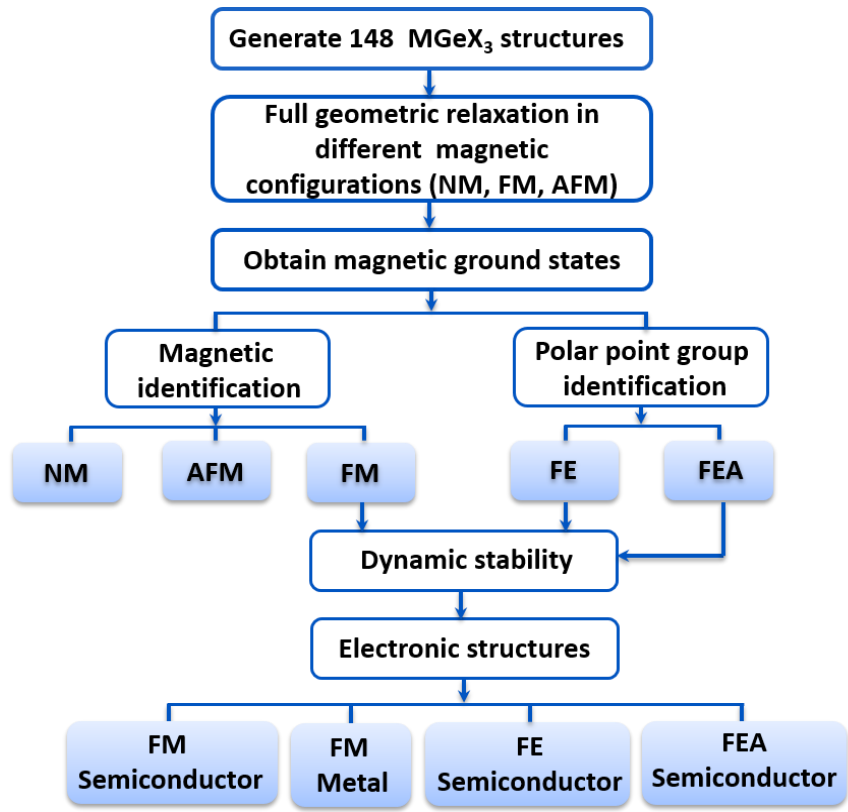

Figure 2. Schematic flowchart of high-throughput first-principles investigations on geometric structures and ferroic properties of 2D MGeX$_3$ family. FM, AFM and NM represent ferromagnetic, antiferromagnetic and non-magnetic, respectively. FE and FEA represent ferroelectric and ferroelastic, respectively.

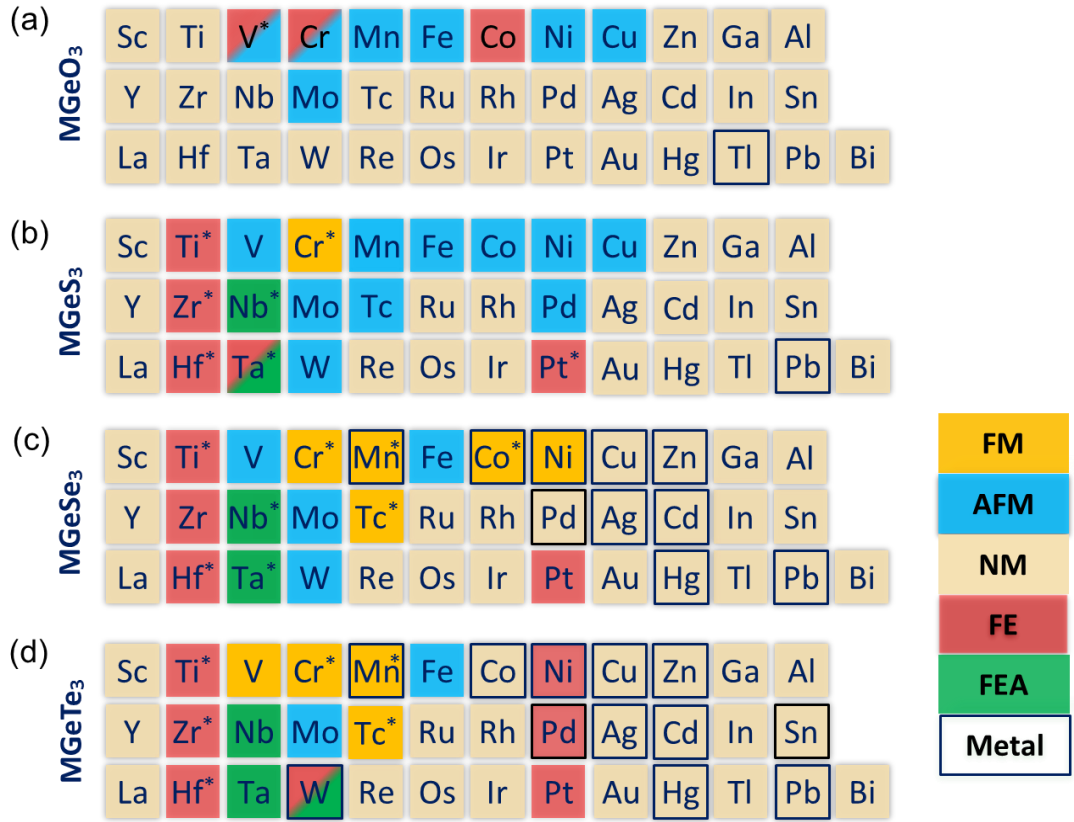

Figure 3. Atlas of ferroicity for 2D MGeX$_3$ (M = metal elements; X = O, S, Se, Te). (a)MGeO$_3$, (b)MGeS$_3$, (c)MGeSe$_3$ and (d)MGeTe$_3$. FM, AFM and NM represent ferromagnetic, antiferromagnetic and non-magnetic, respectively. FE and FEA represent ferroelectric and ferroelastic, respectively. These ferroic properties are indicated by different colors in the block corresponding to each MGeX$_3$ material, and two different colors in the same block implies a potential multiferroic material. The FM and FE materials with verified dynamic stability are marked with asterisks. Metallic materials are indicated with black solid frames.

**3.3 Magnetic properties of 2D MGeX$_3$ family.** As indicated in Fig. 3, among 148 possible 2D MGeX$_3$ materials, ten of them have FM ground states and 21 have AFM ground states, while others are non-magnetic. Eight 2D FM MGeX$_3$ materials are verified to be dynamically stable, including five FM semiconductors, i.e., CrGeS$_3$, CrGeSe$_3$, CrGeTe$_3$, TcGeSe$_3$ and TcGeTe$_3$, and three FM half-metals, i.e., MnGeSe$_3$, MnGeTe$_3$, and CoGeSe$_3$. Note that CrGeTe$_3$ is actually the experimentally discovered 2D magnetic material [1], and the results of other four FM semiconductors are also consistent with previous works [17-19]. Then we focused on 2D FM MnGeSe$_3$, MnGeTe$_3$, and CoGeSe$_3$. Figures 4(a), 4(c), and 4(e) show their electronic energy bands at the GGA+U level, which indicate that they are half-metals with only one species of electrons, say, the spin-up electrons at the Fermi level, while the spin-down energy bands have indirect gaps of 1.67, 0.92 and 1.61 eV, respectively. Interestingly, for 2D MnGeSe$_3$, MnGeTe$_3$, and CoGeSe$_3$, the pairs of band-inversion Weyl points and a band-crossing Weyl point near the Fermi level can be observed along the high symmetry $k$-line Γ-M/Γ-K and M-K, respectively. Thus, they are Weyl half-metals. Considering the spin orbital coupling (SOC) effect, the electronic energy bands at the level of GGA+U+SOC were also calculated as shown in Figs. 4(b), 4(d) and 4(f), from which one may see that tiny gaps open at Weyl points. The SOC gaps of MnGeTe$_3$ are larger than that of MnGeSe$_3$ and CoGeSe$_3$, revealing that a strong spin orbital coupling appears in MnGeTe$_3$. Moreover, the effects of the on-site Coulomb interaction $U$ on magnetic and electronic properties were also studied. The results show that the FM ground state (Fig. 4(g)) and half metallic property (See details in Figs. S3 and S4 in the ESM) for 2D MnGeSe$_3$ and MnGeTe$_3$ are always maintained under different $U$ values. For CoGeSe$_3$, the FM ground state does not change with $U$ = 3, 4, 4.5 eV and the half metallic feature is maintained with $U$ = 4, 4.5 eV (Fig. S5 in the ESM). The electronic structures were reexamined by using HSE06 method (See Fig. S6 in the ESM). The results on 2D MnGeSe$_3$ and MnGeTe$_3$ are consistent with the above GGA+U results, i.e., they are FM half metals. However, the result on 2D CoGeSe$_3$ is different from GGA+U results, indicating it is a nonmagnetic semiconductor. The similar situation that HSE gives different results from GGA+U had also been observed in previous work [19]. Thus, the Coulomb $U$ may play a crucial role in the electronic and magnetic state of 2D CoGeSe$_3$, and further investigation may be necessary.

The total energies of 2D MnGeSe$_3$, MnGeTe$_3$ and CoGeSe$_3$ with different magnetic configurations were obtained at the GGA+U+SOC level (Table 1). The results reveal that the in-plane FM configuration along $y$ direction (FM$^y$) has the lowest energy, and the magnetic anisotropy energy (MAE), defined as the energy difference between the in-plane (FM$^y$) and out-of-plane (FM$^z$) magnetic configuration, i.e., MAE = $E_{FMy}$ - $E_{FMz}$, which are 6.96, 23.49 and 7.13 meV for MnGeSe$_3$, MnGeTe$_3$ and CoGeSe$_3$, respectively.

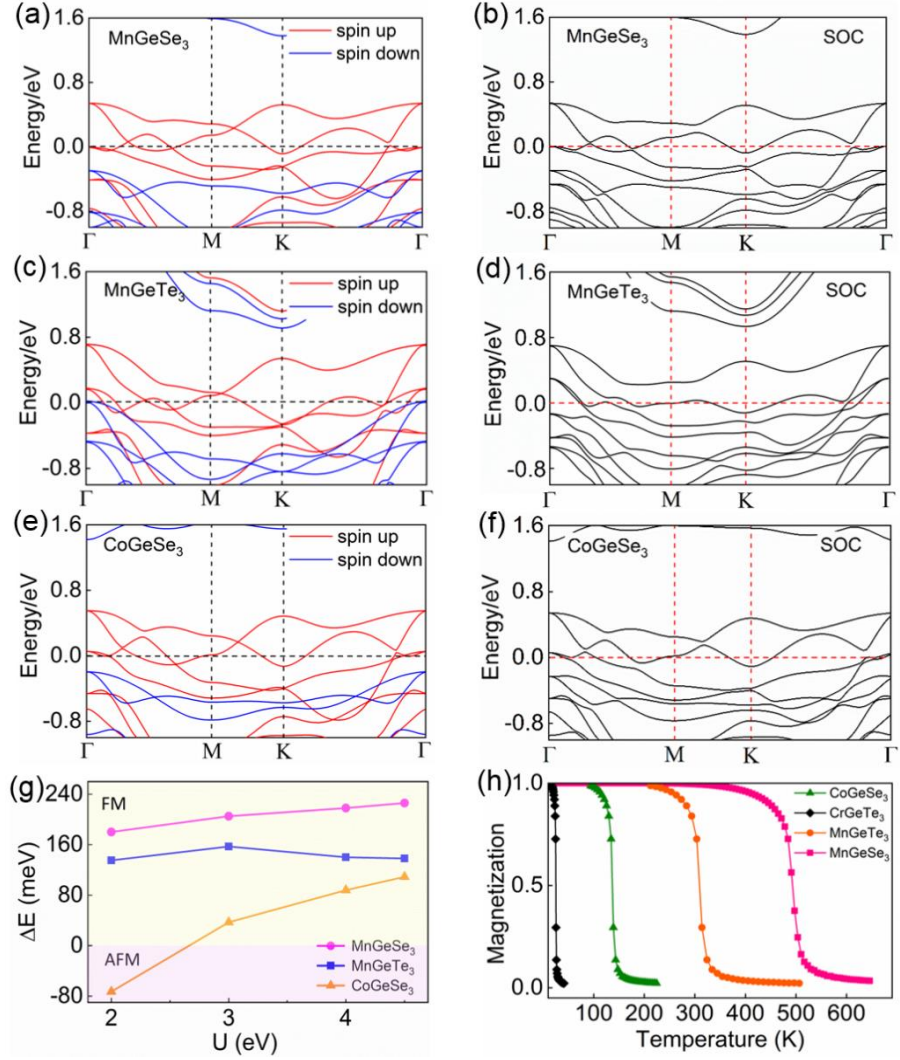

Figure 4. The electronic band structures at the GGA+U and GGA+U+SOC levels of (a)(b) MnGeSe$_3$, (c)(d) MnGeTe$_3$ and (e)(f) CoGeSe$_3$, respectively. (g) Energy differences between the FM state and AFM state (E$_{AFM}$ - E$_{FM}$) as a function of U for 2D MnGeSe$_3$, MnGeTe$_3$, CoGeSe$_3$ and CrGeTe$_3$, respectively. (h) Monte Carlo simulated temperature dependence of the normalized magnetic moment of 2D MnGeSe$_3$, MnGeTe$_3$, CoGeSe$_3$ and CrGeTe$_3$.

Obviously, the MAE of 2D MnGeTe$_3$ is much larger than that of MnGeSe$_3$ and CoGeSe$_3$, which may also reflect a strong SOC in MnGeTe$_3$, since the MAE caused by the single-ion anisotropy is proportional to the square of SOC strength [60]. To describe the magnetic interactions in 2D MnGeSe$_3$, MnGeTe$_3$, and CoGeSe$_3$, the Ising Hamiltonian $H_{spin} = -\sum_{<i,j>} J S_i^y S_j^y$ was adopted, where $J$ represents the nearest-neighbor exchange integral, $S_{i,j}^y$ is the y-component of spin operator, and $<i,j>$ denotes the summation over the nearest neighbors. $J$ could be determined by the energy difference between FM$^y$ configuration and the lowest-energy AFM configuration, which describes the super-exchange interactions that are derived from the hybridization between $p$ orbital of X atoms and $d$ orbital of M atoms. Based on above Hamiltonian, Monte Carlo simulations [61] were performed on an 80×80 honeycomb lattice with $10^6$ step iterations for each temperature to investigate the temperature dependence of magnetization. As shown in Fig. 4(g), the normalized magnetic moments decrease rapidly and ferromagnetic-paramagnetic phase transitions are observed at Curie temperature of 490, 308, 119 K for 2D MnGeSe$_3$, MnGeTe$_3$, and CoGeSe$_3$, respectively. For comparison, the Curie temperature of CrGeTe$_3$ was obtained to be 25K with the same method, which is consistent with the experimental observation [1]. Thus, 2D MnGeSe$_3$ and MnGeTe$_3$ are potential room-temperature 2D FM half-metals.

Table 1. The total energy (meV) per unit cell of MnGeSe$_3$, MnGeTe$_3$, and CoGeSe$_3$ with different magnetic configurations (Figure 1(e)) at the GGA+U+SOC level with respect to the total energy of their ground states (in-plane FM along $y$ direction, FM$^y$). The magnetic moment M ($\mu_B$) and the predicted Curie temperature $T_C$ (K) are also listed.

|  | FM$^x$ | FM$^y$ | FM$^z$ | N-AFM | s-AFM | z-AFM | M($\mu_B$) | $T_C$(K) |
|---|---|---|---|---|---|---|---|---|
| MnGeSe$_3$ | 0.03 | 0 | 6.96 | 244.2 | 208.7 | 356.9 | 4.4 | 490 |
| MnGeTe$_3$ | 0.17 | 0 | 23.49 | 186.6 | 131.6 | 1364.2 | 4.4 | 308 |
| CoGeSe$_3$ | 0.08 | 0 | 7.13 | 72.9 | 191.2 | 321.1 | 2.4 | 119 |

Several interesting points for magnetic properties of 2D MGeX$_3$ family can be observed from Fig. 3 in order. (i) 2D ferromagnetic MGeX$_3$ are sparse, and Cr, Mn, Co and Tc elements are inclined to form ferromagnets; (ii) for MGeO$_3$, there is no intrinsic FM members, and for MGeS$_3$, only one FM member, i.e., CrGeS$_3$; in contrast, there are many AFM MGeO$_3$ and MGeS$_3$ members especially when M = 3$d$ metal elements; (iii) there are four and three stable FM members for MGeSe$_3$ and MGeTe$_3$, respectively, indicating that Se and Te are favored more than O and S to be ferromagnetic in 2D MGeX$_3$; (iv) for specific metallic M atoms in 2D MGeX$_3$, such as Mn, Co, and Tc, transitions from AFM to FM may occur when X atom varies from O and S to Se and Te, which may be due to the exchange interactions between neighboring metal atoms varying with the distance between neighboring metal atoms, which are affected by different sizes of chalcogen atoms. As discussed in previous works on magnetic materials with similar structure [14, 17, 19], the M-X-M bond angles in 2D MGeX$_3$ are nearly 90°, implying the long-range ferromagnetic superexchange interaction exists between two neighboring M atoms mediated by the middle X atoms. Their magnetic ground states may be attributed to the competition effect of direct antiferromagnetic exchange interaction and indirect ferromagnetic superexchange interaction, which is sensitive to various factors including strain [62], doping [63], electrostatic gating [64], *etc.*, exhibiting abundant modulation possibilities of the magnetic properties of 2D $MGeX_3$ family.

**3.4 Ferroelectric properties of 2D $MGeX_3$ family.** Figure 3 also indicates 11 ferroelectric members of 2D $MGeX_3$ family, in which the Ge-Ge pair unexpectedly displaces along the *z* axis while the hexagonal lattice of M atoms holds its original configuration, as shown in Fig. 1(c). Thus, they are displacive-type ferroelectrics, and the physical origin of ferroelectricity comes from the spontaneous symmetry breaking that is mainly induced by vertical displacements of Ge-Ge pairs. Table 2 lists 11 stable 2D ferroelectric $MGeX_3$ materials and their space groups, band gaps, ferroic properties, polarization, energy barriers and piezoelectric coefficients. A few interesting facts can be observed. (i) Most of them contain the M atoms that belong to the IVB group in the periodic table, i.e., Ti, Zr, and Hf; (ii) for 2D ferroelectric $MGeX_3$ with the same M atoms, larger X atoms lead to larger Ge-Ge bond lengths (*d*) and off-centering displacement (*Δd*), while lead to smaller energy band gaps and polarizations; (iii) the Ge-Ge bond lengths of 2D $MGeX_3$ (M = Ti, Zr, and Hf, which are all elements of group IVB) are between 2.60 and 2.70 Å, while the Ge-Ge bond length of $PtGeS_3$ (Pt is the element of group VIII) is 2.48 Å and that of $VGeO_3$ and $TaGeS_3$ (V and Ta are elements of group VB) are 2.78 and 2.81 Å, indicating the Ge-Ge bond length is much affected by the valence electrons of metallic M atoms.

It is worth to mention that 2D $TiGeTe_3$, $ZrGeTe_3$, $HfGeSe_3$, $HfGeTe_3$, $PtGeS_3$ and $VGeO_3$ have remarkable piezoelectric coefficients, which are larger than that of 2D monolayer $MoS_2$ (3.13 pm/V, which is estimated using the same method and is consistent with the previous work [65]) and may have possible applications in pressure sensor or energy conversion devices. Figures 5(a) and 5(b) illustrate the ferroelectric-paraelectric transition pathways of 2D $TiGeTe_3$ and $ZrGeTe_3$, which exhibit typical ferroelectric bistable characters. The energy band structures at the HSE06 level are depicted in Figs. 5(c) and 5(d), implying they are indirect semiconductors with band gap of 0.54 and 0.72 eV, respectively. The density of states for TiGeTe$_3$ was also presented in Fig. 5(e), illustrating that there is an obvious hybridization between the *p*-orbital of Te atoms and *d*-orbital of Ti atoms. The ferroelectric-paraelectric transition pathways and electronic structures of other stable 2D FE MGeX$_3$ materials are also investigated, as shown in Figs. S8 and S9 in the ESM.

Table 2. The space group, band gap (eV) at HSE06 level, geometric parameters, ferroic properties, polarization, energy barriers and piezoelectric coefficients of 11 stable 2D ferroelectric MGeX$_3$ materials. $d_{Ge-Ge}$ (Å) is the length of Ge-Ge bond; $\Delta d$ (Å) is the off-centering displacement between Ge-Ge pair and the hexagonal plane of metal M atoms; P is the spontaneous out-of-plane electric polarization; E$_a$ (eV) is the transition barrier between ferroelectric phase and paraelectric phase; P$_{11}$ (pm/V) is piezoelectric coefficient.

| Material | Space group | Gap (eV) | $d_{Ge-Ge}$ (Å) | $\Delta d$ (Å) | Ferroic | P(pC/m$^2$) (eÅ/unit cell) | E$_a$ (eV) | P$_{11}$ (pm/V) |
|---|---|---|---|---|---|---|---|---|
| TiGeS$_3$ | P31m | 1.19 | 2.62 | 0.867 | FE | 5.187(0.112) | 0.83 | 1.01 |
| TiGeSe$_3$ | P31m | 1.00 | 2.62 | 0.892 | FE | 3.116(0.085) | 0.63 | 0.89 |
| TiGeTe$_3$ | P31m | 0.54 | 2.64 | 0.924 | FE | 1.488(0.047) | 0.32 | 74.06 |
| ZrGeS$_3$ | P31m | 1.86 | 2.63 | 0.882 | FE | 3.438(0.093) | 1.09 | 1.79 |
| ZrGeTe$_3$ | P31m | 0.72 | 2.67 | 0.956 | FE | 1.621(0.054) | 0.33 | 7.52 |
| HfGeS$_3$ | P31m | 2.03 | 2.63 | 0.889 | FE | 3.400(0.090) | 1.20 | 1.31 |
| HfGeSe$_3$ | P31m | 1.64 | 2.66 | 0.943 | FE | 2.785(0.080) | 0.88 | 7.25 |
| HfGeTe$_3$ | P31m | 0.89 | 2.70 | 0.988 | FE | 1.519(0.050) | 0.44 | 15.41 |
| PtGeS$_3$ | P31m | 1.53 | 2.48 | 0.765 | FE | 1.823(0.044) | 0.65 | 11.89 |
| VGeO$_3$ | P31m | 3.06 | 2.78 | 0.886 | FE&AFM | 1.160(0.019) | 3.15 | 9.16 |
| TaGeS$_3$ | Cm | 1.34 | 2.81 | 0.996 | FE&FEA | 6.008(0.151) | 0.60 | 0.41 |

We performed a qualitative statistical analysis on the correlation between polarization and various element-related properties of 2D FE MGeX$_3$. Inspired by the results from a compressed-sensing method, i.e., sure independence screening and sparsifying operator (SISSO) [66], we discover that the polarization (P) is generally in proportion to the ratio of the electronegativity [67] of X atom (E$_X$) and that of M atom (E$_M$), which is shown in Fig. 5(f), where the fitted equation can be obtained as P = $0.184 E_X/E_M - 0.294$. The electronegativity describes the ability of an atom gaining electrons from other atoms, and a large difference between $E_X$ and $E_M$ should induce a strong polar bonding between X and M atoms, which may then lead to the spontaneous symmetry breaking and a high polarization of the whole system. Besides, by noting that no stable 2D FE MGeX$_3$ with $E_X/E_M < 1.6$ was observed, we speculate that a strong polar bonding may be more beneficial to maintain a stable spontaneous symmetry breaking phase in 2D MGeX$_3$ family.

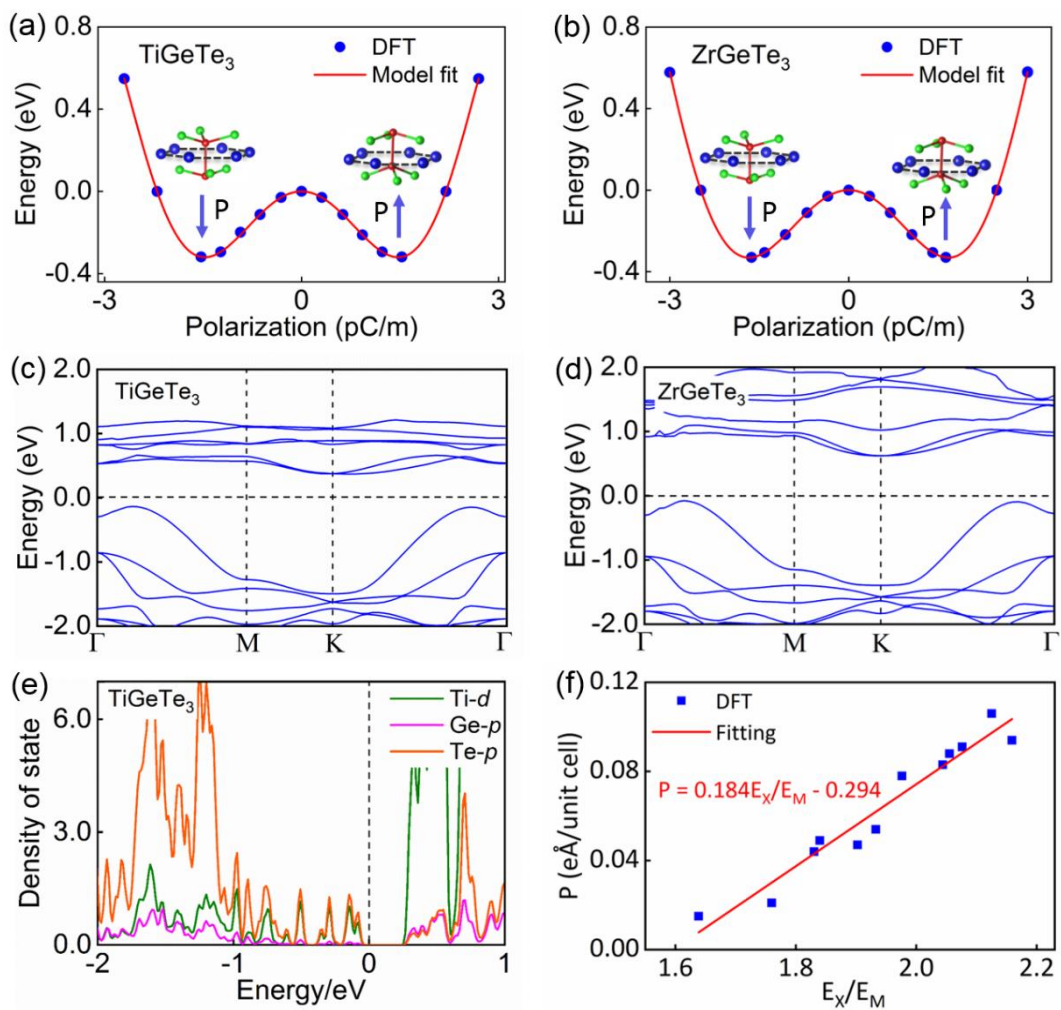

Figure 5. Energy versus polarization of 2D ferroelectric (a) TiGeTe$_3$ and (b) ZrGeTe$_3$. The blue dots represent the data from DFT calculations and the red lines are model fitting lines. The minima correspond to the ferroelectric phases with opposite polarization directions indicated by blue arrows. The band structures at *HSE06* level of 2D (c) TiGeTe$_3$ and (d) ZrGeTe$_3$. (e) The projected electronic density of states of 2D TiGeTe$_3$. (d) The relationship of P (polarization) and $E_X/E_M$ ($E_X$ and $E_M$ is the electronegativity of X atom and M atom, respectively), where the blue dots represent the data from DFT calculations and the red line is model fitting line.

**3.5 Ferroelastic properties of 2D MGeX$_3$ family**. As depicted in Fig. 3, several 2D

ferroelastic MGeX$_3$ members, i.e., TaGeS$_3$, TaGeSe$_3$, NbGeS$_3$, and NbGeSe$_3$ were discovered, and their dynamic stabilities were also verified by phonon dispersions, as shown in Fig. S10 in the ESM. All of them contain M atoms that belong to VB group, i.e., Nb and Ta. The schematic geometric structures are shown in Fig. 1(d), including three equivalent phases $\alpha$, $\beta$ and $\gamma$, in which the Ge-Ge pairs are inclined to three symmetric directions, respectively. The three equivalent phases could convert mutually with switching barriers of 0.38, 0.54, 0.51, 0.57 eV for TaGeS$_3$, TaGeSe$_3$, NbGeS$_3$, and NbGeSe$_3$, respectively. The transformation paths between different configurations are investigated, as illustrated in Fig. S11 in the ESM. These 2D ferroelastic MGeX$_3$ materials may have potential applications in shape memory devices.

**3.6 Multiferroics in 2D MGeX$_3$ family.** In Fig. 3, two different colors in the same block imply a potential multiferroic material that involves two different types of ferroic orderings. 2D VGeO$_3$ is a multiferroic material with ferroelectricity and antiferromagnetism (FE & AFM). As indicated in Table 2, 2D VGeO$_3$ has the same space group with other 2D FE members and the largest energy gap of 3.06 eV. It has a Néel AFM ground state and the magnetic momentum of V atom is 1.9 $\mu_B$, which mainly comes from $3d$ orbitals. 2D TaGeS$_3$ is another multiferroic material with ferroelectricity and ferroelasticity (FE & FEA), in which Ge-Ge pairs have simultaneous vertical and tilted displacements along three equivalent directions (See Fig. 1(d)). The FE and FEA properties of 2D TaGeS$_3$ are presented in Table 2 and Fig. S11 in the ESM. The coexistence of different ferroic orderings in the same materials provides various modulation possibilities, which may be applied to construct multifunctional devices.

**3.7 Potential Applications of 2D ferroic MGeX$_3$ materials**

**Magnetic tunnel junctions based on 2D MGeX$_3$ half-metals.** Half metals are conducting in only one spin channel while the other spin channel is insulator or semiconductor. They are ideal materials for spintronic devices [68, 69], as they could provide fully spin-polarized currents and maximize the magnetoresistance in spin valves and magnetic tunnel junctions. Here, a magnetic tunnel junction based on 2D MGeX$_3$ half-metals is proposed, in which the vertical heterostructures combine 2D half-metallic MnGeSe$_3$ and MnGeTe$_3$ with in-plane magnetizations and a separating insulating tunneling layer, as depicted in Fig. 6(a). The magnetization of the top layer (MnGeSe$_3$) can be adjusted by an external magnetic field, while the magnetization of bottom layer (MnGeTe$_3$) can be pinned by a ferromagnetic substrate. When the magnetization of the top layer is switched from parallel to antiparallel to that of the bottom layer, the whole magnetic tunnel junction will change from low-resistance state to high-resistance state, and thus the tunnel magnetoresistance (TMR) effect can be expected [5, 70]. Besides, these 2D MGeX$_3$ half metals may also have great potential applications in ultrathin spin filter and spintronic transistor.

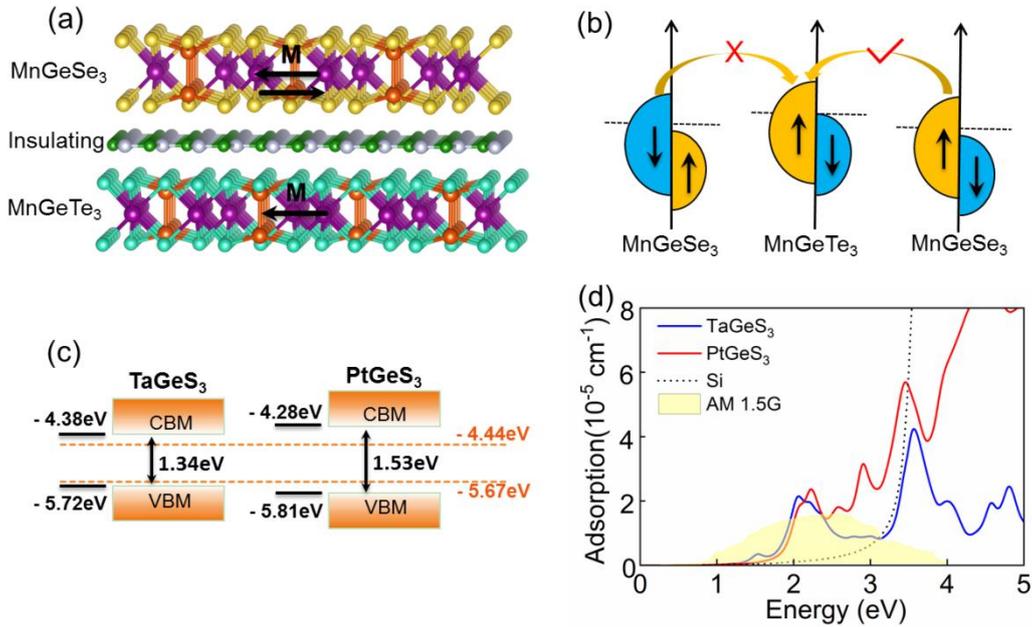

Figure 6. (a) Magnetic tunnel junction based on 2D half-metallic MnGeSe$_3$ and MnGeTe$_3$ with an insulating layer in the middle. The magnetization of the top layer can be adjusted by an external magnetic field, while the magnetization of bottom layer can be pinned by a ferromagnetic substrate. (b) Schematic depiction of the tunneling between MnGeSe$_3$ and MnGeTe$_3$ in the magnetic tunnel junction when magnetization of MnGeSe$_3$ is reversed. (c) The band gap and band edges of 2D ferroelectric TaGeS$_3$ and PtGeS$_3$ with respect to the vacuum level, where the water redox potential (between -4.44 eV and -5.67 eV) is also included for comparison. CBM and VBM represent conduction band minimum and valence band maximum, respectively. (d) The absorption spectra of 2D ferroelectric TaGeS$_3$ and PtGeS$_3$ at the G$_0$W$_0$+BSE level. The adsorption spectra of intrinsic silicon (Si) is included for comparison. The yellow area indicates the incident AM1.5G solar flux.

**2D ferroelectric MGeX$_3$ photocatalyst for splitting water.** Ferroelectricity could benefit photocatalytic water-splitting performance because the built-in electric field can effectively hinder the recombination of photogenerated electrons and holes, and thus boost the solar-to-hydrogen efficiency [47, 71]. After reexamining the electronic structures of all 2D FE MGeX$_3$ members (Fig. S9 in the ESM), we found that 2D ZrGeS$_3$, HfGeS$_3$, HfGeSe$_3$, TaGeS$_3$ and PtGeS$_3$ have not only appropriate band gaps (1.2 eV ~ 3.0 eV at HSE06 level) but also band edges perfectly straddling the water redox potential; i.e., the conduction band minimum (CBM) lies above the reduction reaction potential (−4.44 eV with respect to the vacuum level) and valence band maximum (VBM) lies below the oxidation reaction potential (−5.67 eV with respect to the vacuum level), as shown in Fig. 6 (c) and Fig. S12 in the ESM. Then the absorption spectra at the $G_0W_0$+BSE level were calculated, which indicate that the optical absorption spectra of 2D TaGeS$_3$ and PtGeS$_3$ cover almost the entire incident solar spectrum with optical absorption peak at 2.073 and 2.234 eV, respectively, as shown in Fig. 6(d). Particularly, 2D PtGeS$_3$ exhibits an excellent optical absorption within the near-infrared light region as well as the ultra-violet region, a proper H$_2$O adsorption, and a rather low overpotential (See Fig. S13 in the ESM), implying a promising 2D ferroelectric photocatalyst for water-splitting.

## 4. CONCLUSIONS

In summary, by means of the high-throughput first-principles calculations, we thoroughly explore the ferroic properties of 2D MGeX$_3$ family and obtain a comprehensive ferroic atlas. We discover eight stable 2D ferromagnetic materials including five FM semiconductors and three half-metals, 21 2D AFM materials, and 11 stable 2D FE semiconductors including two multiferroic materials. We found that Cr, Mn, Co and Tc elements are beneficial to form 2D FM MGeX$_3$, while Se and Te are much favored than O and S for FM. Particularly, 2D CoGeSe$_3$, MnGeSe$_3$ and MnGeTe$_3$ are predicted to be high temperature FM half metals with Curie temperature of 119, 490 and 308 K, respectively. It is probably for the first time that ferroelectricity is discovered in 2D MGeX$_3$ family, which derives from the spontaneous symmetry breaking induced by the unexpected displacement of Ge-Ge atomic pairs. We also reveal that the polarizations are in proportion to the ratio of electronegativity of X and M atoms, and metal elements of IVB group are highly favored for ferroelectricity, while metal elements of VB group are favoured for ferroelasticity in 2D $MGeX_3$. Furthermore, magnetic tunnel junctions based on 2D FM $MGeX_3$ half-metals and water-splitting photocatalysts based on 2D FE $MGeX_3$ are proposed as examples of their wide applications. The atlas of ferroicity in 2D $MGeX_3$ materials we obtained here highly enriches the family of 2D ferroic materials and will motivate great interest in exploring 2D $MGeX_3$ family as novel potential multi-functional materials.

## CONFLICTS OF INTEREST

There are no conflicts to declare.

## ACKNOWLEDGEMENTS

The authors would like to thank Prof. Bo Gu, Prof. Zheng-Chuan Wang, Dr. Jing-Yang You, and Ms. Zhen Zhang for helpful discussions. All calculations are performed on Tianhe-2 at National Supercomputing Center in Guangzhou, China. This work is supported in part by the National Key R&D Program of China (Grant No. 2018YFA0305800), the Strategic Priority Research Program of Chinese Academy of Sciences (Grant No. XDB28000000), the National Natural Science Foundation of China (Grant No. 11834014), the Beijing Municipal Science and Technology Commission (Grant No. Z181100004218001), the fundamental research funds for the central universities and University of Chinese Academy of Sciences.

Electronic Supplementary Material: Supplementary material (The phonon dispersions, electronic structures on different levels, and molecular dynamics simulations of 2D ferroic $MGeX_3$ family; the phase transitions of 2D ferroelectric and ferroelastic $MGeX_3$ materials.) is available in the online version of this article at https://doi.org/10.1007/s12274-021-3415-6.